\documentstyle[12pt,epsf]{procsla}
\input{epsf.sty}
\begin{document}

\begin{center}
\begin{Large}
\begin{bf}
 CHARMED QUARK AND $J/\Psi$ PHOTOPRODUCTION
 IN THE SEMIHARD APPROACH OF QCD
  AT HERA ENERGIES
\end{bf}
\end{Large}
\vspace*{1cm}

    V.A.Saleev\\    
{\it Samara State University, Samara 443011, Russia\\}

\vspace*{0.5cm}
    N.P.Zotov\\ 
{\it D.V. Skobeltsyn Institute of Nuclear Physics,Moscow State University,\\ 
      Moscow 119899, Russia\\}
\end{center}      
\vspace*{1.5cm}
\begin{abstracts}
{\small We compare our theoretical results for $c\bar c-$quark and
$J/\Psi-$photoproduction cross sections obtained in the semihard approach
 of QCD~\cite{r1,r2} with H1 and ZEUS experimental data~\cite{r3,r4,r5}} 
\end{abstracts}

\vspace*{1.5cm} 
\section{Introduction}
The observation of heavy quark and quakonium photoproduction offers a
unique opportunity to probe the gluon distribution in proton at small $x$ 
by measurements of the total cross sections of these processes or of their
differential distributions.
 
 It is known that at HERA energies and beyond heavy quark and
quarkonium photoproduction processes
 are of the so-called semihard type~\cite{r6}.
In such 
processes by definition a hard scattering scale $Q$ (or heavy quark
mass $M$) is large as compared to the $\Lambda_{QCD}$ parameter,
but $Q$ is much smaller than the total center of mass energies:
$\Lambda_{QCD} \ll Q \ll \sqrt s$. The last condition implies that 
the processes occur in small $x$
region: $x\simeq M^2/s\ll 1$. In such a case the perturbative
QCD expansion has large coefficients $O(\ln^{n}(s/M^2))
\sim O(\ln
^{n}(1/x))$ besides the usual renormalization group ones
which are $O(\ln^{n}(Q^2/\Lambda^2))$.
It means that in the small $x$ region $(x\simeq M^2/s \ll 1)$ we should 
first of all sum up all the terms of the type $(\alpha_s\ln s/M^2)^n$ and 
also the terms $(\alpha_s\ln Q^2/Q^2_0\ln s/M^2)^n$ (moreover $\alpha_s\ln Q^2/Q^2_0$
in DIS). The second problem that appears at small $x$ is that of the violation of
unitarity due to the increase of the gluon density at $x\to 0$. The latter one
is reduced to taking into account the absorption (screening) corrections which should 
stop
the growth of the cross section at $x\to 0$ (in accordance with the unitarity
condition).
Both these problems were resolved in the so called semihard approach (SHA) by 
L. Gribov, E. Levin, M. Ryskin~\cite{r6}.
Therefore we used this SHA for the calculations
of the cross sections of $J/\Psi -$ and $c\bar c -$photoproduction~\cite{r1,r2}.
In present paper the result of these calculations will be compared with new H1
and ZEUS experimental data at HERA enargies~\cite{r3,r4,r5}.

\newpage
\section{Heavy quark photoproduction cross section in the semihard
approach}
In the SHA heavy quark photoproduction is determined by the contribution
of two photon-gluon fusion diagrams
with "unusual" properties of the gluons in proton. These gluons are the off mass shell
ones, their virtuality $q^2 = -\vec{q}_T ^2$, and they possess the propeperty of
 the alignment of their polarization vector along their transverse momentum
such as $\epsilon_{\mu} = q_{T\mu}/|\vec{q}_{T}|$~\cite{r6,r7}.
Their distributions in $x$ and $q_T$ in proton is given by
the unintegrated gluon structure function $\varphi (x,q^2_T)$,
 wich is connected with the usual gluon 
distribution function $G(x,Q^2)$ by the following relation
\begin {eqnarray}
\int\limits_0^{Q^2}\varphi
(x,q_{T}^2)dq_{T}^2=xG(x,Q^2).
\end{eqnarray}
The exact expression for the function $\varphi (x,q^2)$ can be obtained as a 
solution of the evolution equation, which, contrary to the parton model case,
is nonlinear due to interactions between the gluons in small $x$ region.

In our calculations we used the following phenomenological parametrization~\cite{r7}:
\begin{eqnarray}
\varphi(x,q_{T}^2)=C\frac{0.05}{x+0.05}(1-x)^3f(x,q_{T}^2),
\end{eqnarray}
where
\begin{eqnarray}
f(x,q^2_T)& =&1 ,\qquad q_{T}^2\le q_0^2(x)\nonumber\\
f(x,q^2_T)& =&(\frac{q_0^2(x)}{q_{T}^2})^2, \qquad q_{T}^2>q_0^2(x),
\end{eqnarray}
and $q^2_{0}(x) = Q^2_{0} + \Lambda^2\exp( 3.56 \sqrt{ \ln(x_0/x)})$,
 $Q_{0}^2 = 2 GeV^2$, $\Lambda = 56$ MeV, $x_{0}$ = 1/3.
The parameter $q_0^2(x)$ can be treated as a new infrared-cutoff, which plays
the role of a typical transverse momentum of partons in the parton cascade
of the proton in semihard processes. The behaviour of $q_0(x)$ was 
discussed in~\cite{r6}. It increases with $\ln (1/x)$ and at $x =0.01 - 0.001$
the values of $q_0(x)$ are about $2 - 4 GeV$.

For relatively small virtuality $Q^2 \le q^2_0(x)$ the gluon function
behaves as $xG(x,Q^2) \sim CQ^2$. So we have the saturation of the gluon density
 at
these values $Q^2$.

 The normalization factor $C\simeq 0.97$ mb
 in (2) was obtained in~\cite{r7} where
$b\bar b$-pair
production at Tevatron energy was described.

Thus, in the case of real transverse polarized photon the heavy quark photoproduction
cross section at $x\ll 1$ can be expressed as ~\cite{r2}
\begin{eqnarray}
\frac{d\sigma}{d^2p_{1T}}(\gamma p\to Q\bar Q X)=
\int dy_1^{\ast}\frac{d^2q_{T}}{\pi}
\frac{\varphi (x,q_{T}^2)|\bar M|^2}{16\pi^2(sx)^2\alpha},
\end{eqnarray}
where $p_T, y^{\ast}_1$ are transverse momentum and rapidity (in the center of mass
frame of colliding particles) of heavy quark and $\alpha = 1 - \alpha _1$ with
$\alpha _1 = M_1\exp (y^{\ast}_1)/\sqrt s$.

The matrix element $\bar M$ for a subprocess $\gamma g^{\ast} \to
c\bar c$ depends on the virtuality of the gluon and differs from the one
of the usual parton model. For the square of this matrix element we used the
following form:
\begin{eqnarray}
|\bar M|^2=16\pi^2e_Q^2\alpha_s\alpha_{em}(xs)^2[\frac{\alpha_1^2+\alpha^2}
 {(\hat t-M^2)(\hat u-M^2)}+\frac{2M^2}{q_{T}^2}
(\frac{\alpha_1}{\hat u-M^2}-
 \frac{\alpha}{\hat t-M^2})^2],
\end{eqnarray}
where $\hat s,~\hat t,~\hat u$ are usual
Mandelstam variables of partonic subprocess $\gamma g^{\ast} \to c\bar c$.

\section{Quarkonium photoproduction cross section in SHA of QCD and in
colour singlet model}
 We also used similar formulas for heavy quarkonium photoproduction~\cite{r1}.
In the case of real transverse polarized photon, the heavy quarkonium
photoproduction cross section at $x\ll 1$ in SHA is expressed in the form
(we consider here $J/\Psi-$ photoproduction):
\begin {eqnarray}
\sigma (\gamma p\to J/\Psi X)=\int \frac{dz}{z(1-z)}\int dp_T^2\int
\frac{d\phi}
{2\pi}\int d\vec q_T\,^2 \frac{\varphi (x,q_T^2)}
{16\pi (s+q_T^2)^2}
 \sum|\bar M(\gamma g^*\to J/\Psi g)|^2,
\end{eqnarray}
 where (in the lab. frame) $z=E/E_{\gamma}$, $s=2m_pE_{\gamma}$ and
 $p=(E,\vec p_T,p_z)$ is 4--momentum of the quarkonium, $\phi$ is
  the angle between initial gluon and quarkonium transverse
  momenta $\vec q_T$ and $\vec p_T$.
  In (6) $\sum$ indicates an average over two photon polarizations
  and colours of initial gluon
  as well as a sum over polarizations of final particles.

If we take the limit of zero $\vec q_T$, and if we average over the
transverse directions of $\vec q_T$, we obtain the formula of standard
parton model (SPM):
\begin {eqnarray}
\sigma (\gamma p\to J/\Psi X)=\int \frac{dz}{z(1-z)}\int dp_T^2
 \frac{xG(x,Q^2)}
{16\pi (sx)^2}
 \sum|M_{part}(\gamma g\to J/\Psi g)|^2,
\end{eqnarray}
where $\sum$ now indicates an average over colours and
transverse polarizations of real initial gluon and photon
  as well as a sum over polarizations of final particles. We averaged
  over the transverse directions of $\vec q_T$ using expression:
  \begin{equation}
 \int d\vec q_T\,^2\int \frac{d\phi}{2\pi}\varphi(x,\vec q_T)
  \sum |\bar M|^2=xG(x,Q^2)\sum|M_{part}|^2,
   \end{equation}
 where
 \begin{equation}
  \int_{0}^{2\pi}\frac{d\phi}{2\pi}\frac{q_{T\mu}q_{T\nu}}{\vec
q_T\,^2}
   =\frac{1}{2}g_{\mu\nu}.
 \end{equation}

Within the framework of perturbative QCD the photoproduction of $J/\Psi$
 particles is described by subprocess $\gamma g\to
  J/\Psi g$~\cite{r8}. In this approach, so called
  "colour singlet model" (CSM), the quarkonium is represented by a define
 wave function so that the final $c\bar c$ system are colour
 singlet, $J^p=1^-$ state of specified mass. At not very large $p_T$ the
 contribution of these subprocesses are significantly greater than others,
 such as: $\gamma g\to b\bar b$ with $b\to J/\Psi X$.
 That is why we took into account only this.

The matrix element $\bar M$ of the process $\gamma g\ast \to J/\Psi g$
 was calculated using the sum of six diagrams according to the
 CSM. We makes summation over spins and colours of final gluon, $J/\Psi$
 and photon as in SPM. In the case of initial off shell gluon with
 transverse momentum $\vec q_T$ we takes polarization tenzor in 
the form~\cite{r6,r7}:
 \begin{equation}
  d_{\mu\nu}(q)=\varepsilon_{\mu}(q)\varepsilon_{\nu}(q)=
  \frac{q_{T\mu}q_{T\nu}}{\vec q_T\,^2}.
   \end{equation}
\newpage
 The calculation of $\sum|\bar M|^2$ was made analytically by "REDUCE"
 system and result can be expressed in the following form:
 \begin{equation}
 \sum|\bar M|^2=\frac{Bx^2}{\vec q_T\,^2}\sum_{i=1}^{6}F_i(z,\vec q_T\,^2,
  \hat t,\hat u),
\end{equation}
 where
 \begin{equation}
  B=\frac{32\pi^2}{3\alpha}\alpha_s(Q^2)\alpha_s(q^2)\Gamma_{ee}M,
 \end{equation}
  $Q^2=M^2+\vec p_T\,^2$, M is the mass of quarkonium, $\Gamma_{ee}$ is
   the leptonic quarkonium width, $\hat t$ and $\hat u$ are ordinary
   Mandelstam variables, $\hat t=M^2-(M^2+\vec p_T\,^2)/z$,
   $ \hat u=M^2-\vec q_T\,^2-zxs+2p_Tq_T\cos \phi$ and $x=(M^2-\hat u
      -\hat t)/s$. The explicit form of functions $F_i$ are given in~\cite{r9}.

 We would like to note the limits of applicability of CSM:
$z \leq 0.8$ and $p^2_T \geq 0.1M^2$~\cite{r10}, where $z =
E_{J/\Psi}/E_{\gamma}$ (in lab. frame) and $p_T$ is the $J/\Psi-$ transverse
momentum. These limits correspond to the region of H1 experimental data~\cite{r4}.

\section{Results}
The results of our calculations for cross sections of
$\gamma p \to c\bar c X$ and $ \gamma p \rightarrow J/\Psi X$  processes~\cite{r1,r2}
 are compared (in Figs. 1, 2) with H1 and ZEUS experimental data~\cite{r3,r4,r5}.
(The curves in Fig.1 correspond to different values of $c-$quark mass:
solid - $m_c$ = 1.3 $GeV$, dashed - 1.4 $GeV$ and dash-dotted - 1.5 $GeV$.
)

\unitlength=1cm

\begin{figure}[ht]
\begin{picture}(16,10)
\epsfysize =12cm
\epsfxsize =14cm
\put(+1,-2){\epsfbox{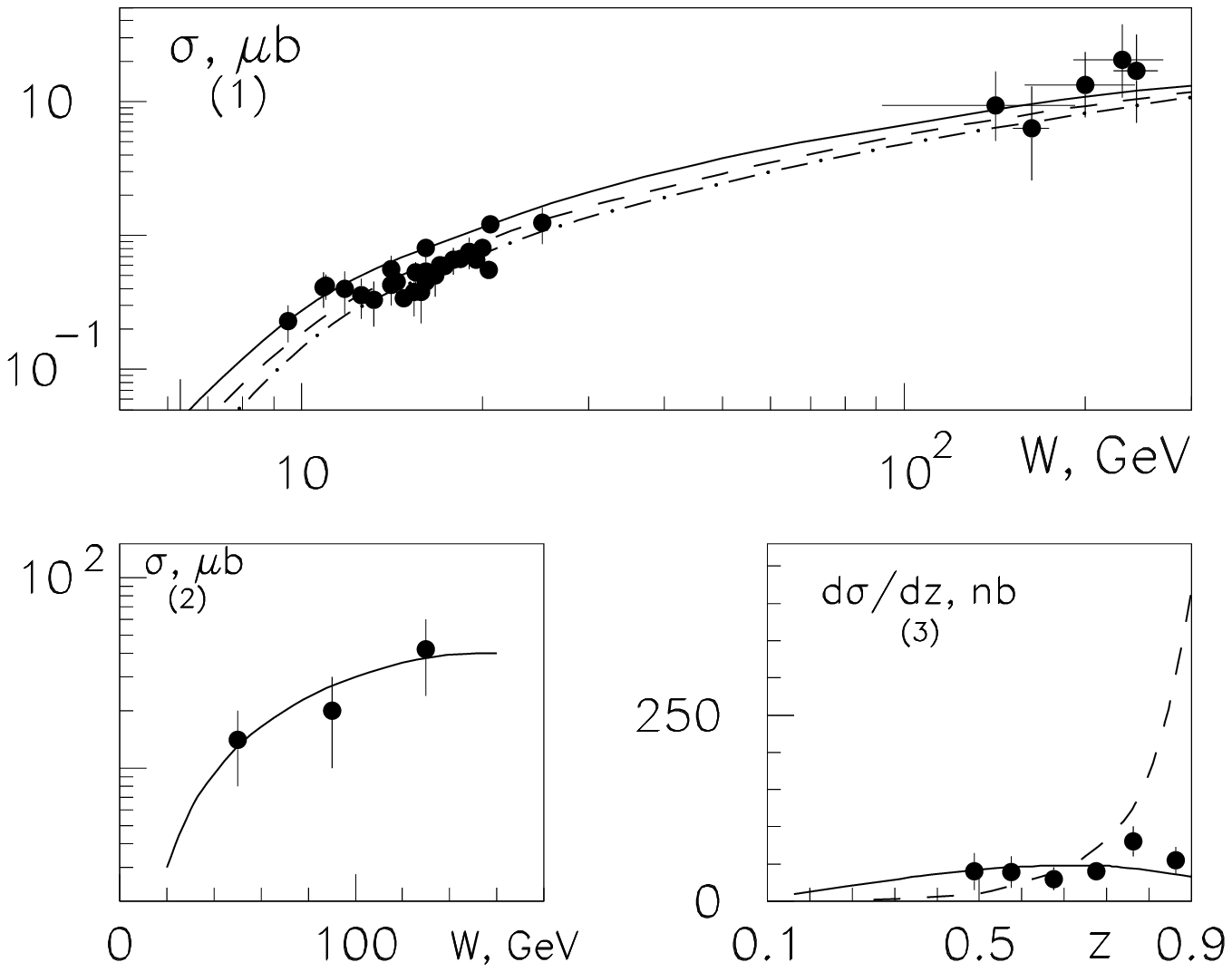}}
\end{picture}
\end{figure}

 In Fig. 3 the z-distribution of $J/\Psi-$mesons~\cite{r1}
is shown (solid curve) in comparison  with H1 experimental data \cite{r4}
and colour octet
model (COM) result (dash curve)~\cite{r11}. 

\newpage
\section{Conclusions}
We see that our theoretical curves obtained in~\cite{r1,r2} describe very
well the new H1 and ZEUS  data, and the strong increase of z-distribution
obtained in  COM is not confirmed by the experimental data~\cite{r4} and our
CSM results~\cite{r1}.

Thus it means that the results obtained in the semihard approach of QCD
can be used for  extraction of the effective gluon distribution in proton from
the H1 and ZEUS experimental data for heavy quark and quarkonium photoproduction
at HERA energies.

\end{document}